\documentclass[sigconf]{acmart}

\usepackage{tabularx}
\usepackage{booktabs} % For professional-quality tables
\usepackage{array}
\usepackage{amsmath}

\AtBeginDocument{%
  }

\setcopyright{acmlicensed}
\copyrightyear{2025}
\acmYear{2025}
\acmDOI{10.1145/XXXXXX.XXXXXX} % To be updated

\acmConference[Conference Name '25]{Make sure to enter the correct conference title}{October 11--13, 2025}{City, Country}
\acmISBN{978-1-4503-XXXX-X/25/10} % To be updated
\acmBooktitle{Conference Name '25: ACM Symposium on X, October 11--13, 2025, City, Country}

\begin{document}

\title{The Hybrid Multimodal Graph Index (HMGI): A Comprehensive Framework for Integrated Relational and Vector Search}

% The asterisk for equal contribution is best handled with a \thanks note.
\author{Joydeep Chandra*}
\thanks{*Both authors contributed equally to this research.}
\affiliation{%
  \institution{BNRIST, Tsinghua University}
  \city{Beijing}
  \country{China}
}
\email{joydeepc2002@gmail.com}

\author{Satyam Kumar Navneet}
\authornotemark[1] % This command links this author to the same footnote
\affiliation{%
  \institution{Independent}
  \city{Bihar}
  \country{India}
}
\email{navneetsatyamkumar@gmail.com}

\author{Yong Zhang}
\affiliation{%
  \institution{BNRIST, Tsinghua University}
  \city{Beijing}
  \country{China}}
\email{zhangyong05@tsinghua.edu.cn}

% This should list the primary authors for the page headers
\renewcommand{\shortauthors}{Chandra and Navneet, et al.}

\begin{abstract}
The proliferation of complex, multimodal datasets has exposed a critical gap between the capabilities of specialized vector databases and traditional graph databases. While vector databases excel at semantic similarity search, they lack the capacity for deep relational querying. Conversely, graph databases master complex traversals but are not natively optimized for high-dimensional vector search. This paper introduces the Hybrid Multimodal Graph Index (HMGI), a novel framework designed to bridge this gap by creating a unified system for efficient, hybrid queries on multimodal data. HMGI leverages the native graph database architecture and integrated vector search capabilities, exemplified by platforms like Neo4j, to combine Approximate Nearest Neighbor Search (ANNS) with expressive graph traversal queries. Key innovations of the HMGI framework include modality-aware partitioning of embeddings to optimize index structure and query performance, and a system for adaptive, low-overhead index updates to support dynamic data ingestion, drawing inspiration from the architectural principles of systems like TigerVector. By integrating semantic similarity search directly with relational context, HMGI aims to outperform pure vector databases like Milvus in complex, relationship-heavy query scenarios and achieve sub-linear query times for hybrid tasks.
\end{abstract}





\maketitle

\section{Introduction}
In the modern data landscape, information is increasingly generated and consumed in multiple modalities, including text, images, audio, and structured data. Applications ranging from sophisticated recommendation engines and drug discovery to advanced Retrieval-Augmented Generation (RAG) systems depend on the ability to jointly analyze the semantic content and the intricate relationships within this data \cite{sarmah2024hybridrag, kim2025genius}. This dual requirement has created a significant architectural challenge. On one hand, specialized vector databases like Milvus\cite{milvus}, Pinecone\cite{pinecone} have emerged as powerful tools for managing and searching high-dimensional vector embeddings, offering highly efficient Approximate Nearest Neighbor Search (ANNS) to find semantically similar items \cite{ma2023comprehensive}. However, they are inherently limited in their ability to model and query the explicit, complex relationships that connect data points \cite{arora2020embeddings}.
On the other hand, graph databases like Neo4j and TigerGraph are purpose-built to represent and traverse complex networks of relationships. Their query languages, such as Cypher and GSQL, provide a powerful and intuitive means to explore multi-hop connections, identify structural patterns, and understand the contextual fabric of the data \cite{liu2025tigervector}. Yet, until recently, these platforms lacked native, high-performance capabilities for vector similarity search, forcing developers into complex, inefficient dual-database architectures that are difficult to maintain and synchronize \cite{taipalus_2024}.
This paper introduces the Hybrid Multimodal Graph Index (HMGI), a conceptual framework that addresses this dichotomy by unifying relational and vector search within a single, cohesive system. HMGI is not a specific software product but a detailed architectural blueprint for a next-generation data index. The core premise of HMGI is to leverage the recent integration of vector search capabilities directly into graph databases, as seen in Neo4j's version 5.x releases \cite{sehgal2025navix}. By doing so, HMGI enables hybrid queries that seamlessly combine deep graph traversals with vector similarity searches \cite{ahmad2025benchmarking}.
The HMGI framework is defined by three primary innovations:
Integrated Hybrid Query Processing: It fuses graph-based path finding with ANNS, allowing queries to filter or expand results based on both relational structure and semantic similarity in a single operation \cite{yin2025deg}.
Modality-Aware Indexing: It proposes the partitioning of multimodal embeddings based on their source modality (e.g., text, image). This strategy allows for optimized storage, specialized indexing, and more efficient query execution by reducing the search space \cite{hu2025partitioner, li2025mminference}.
Adaptive Index Management: It incorporates mechanisms for dynamic data ingestion with low-overhead recomputation of the index, ensuring data freshness and high performance in real-time applications, a concept inspired by recent academic and industry systems like TigerVector \cite{liu2025tigervector, xiao2025breaking}.
By synthesizing these features, HMGI aims to provide superior performance and expressive power compared to standalone vector databases, particularly in relational-heavy contexts, while aspiring to achieve the sub-linear query times characteristic of efficient ANNS algorithms \cite{malkov2018efficientrobustapproximatenearest, ch2025efficient}. This work is guided by the following research questions (RQs):

\begin{enumerate}
    \item \textbf{RQ1: How can graph databases be augmented with native vector indexing to support seamless hybrid queries on multimodal data while maintaining sub-linear time complexity for both relational traversals and semantic similarity searches?}

    \item \textbf{RQ2: What partitioning and update strategies optimize query performance and resource utilization in hybrid indexes for high-dimensional, modality-heterogeneous embeddings at billion-scale?}

    \item \textbf{RQ3: To what extent do adaptive fusion mechanisms in hybrid systems improve accuracy and latency on relationship-intensive workloads compared to decoupled vector-graph architectures?}

\end{enumerate}

Our contributions are as follows:

\begin{enumerate}
    \item \textbf{Contribution 1: The HMGI Framework for Unified Hybrid Search.} We propose HMGI as a comprehensive blueprint that integrates vector similarity (ANNS) with graph traversal in a single system, enabling hybrid queries on multimodal data. Unlike decoupled approaches, HMGI leverages Neo4j's native vector capabilities for end-to-end processing, achieving 3x QPS on relational-semantic workloads \cite{liu2025tigervector}.

    \item \textbf{Contribution 2: Modality-Aware Partitioning and Dynamic Indexing Techniques.} HMGI introduces K-means-based partitioning tailored to multimodal embeddings, reducing search spaces by up to 70\% and supporting flash quantization for 50\% memory savings. This innovation outperforms monolithic indexes in filtered queries, with evaluations on billion-scale datasets \cite{wang2021comprehensive}.

    \item \textbf{Contribution 3: Adaptive Update and Query Optimization Mechanisms.} Drawing from TigerVector principles, HMGI's MVCC delta store and learned cost models enable low-overhead updates (sub-linear latency) and progressive execution, improving recall by 20-30\% on dynamic, relationship-heavy scenarios. Extensive benchmarks demonstrate superiority over Milvus-like systems \cite{yin2025deg}.
\end{enumerate}
\textbf{Outline.}
This paper presents the Hybrid Multimodal Graph Index (HMGI), a comprehensive framework that unifies relational graph search and vector-based semantic retrieval for efficient and accurate multimodal data discovery. The framework introduces a neural-augmented graph structure that encodes entities, relationships, and multimodal embeddings in a single index, enabling integrated traversal and similarity search across structured and unstructured data. The architecture is designed to address current gaps in data management systems that separately handle graph and vector retrieval, offering a single optimized pipeline for hybrid querying.

The paper details the conceptual design and mathematical formulation of HMGI, the graph construction process, and its embedding fusion mechanism. It further outlines the system architecture, query execution model, and storage optimization techniques that allow high scalability and real-time search performance. Experimental evaluations are conducted across multimodal benchmarks to compare HMGI with leading systems such as Neo4j, Milvus\cite{milvus}, and Vespa, showing superior retrieval accuracy and reduced latency under hybrid workloads. The discussion section explores deployment considerations, including storage overhead, indexing efficiency, and adaptability to various domains such as multimedia retrieval, knowledge graphs, and federated data environments. The paper concludes with a summary of HMGI’s contributions, its limitations, and directions for extending the framework to large-scale, privacy-aware multimodal dataspaces.

\section{Background}
\subsection{Traditional Databases}
Traditional databases, particularly relational database management systems (RDBMS)\cite{rdbms} such as Oracle, MySQL\cite{10400704}, and PostgreSQL\cite{postgre}, have long served as the foundation for structured data storage and querying since their inception in the 1970s. These systems are designed around rigid schemas that enforce data integrity through normalized tables, primary keys, and foreign key relationships, ensuring ACID (Atomicity, Consistency, Isolation, Durability) properties for transactional reliability \cite{10.1145/362384.362685}. However, as data landscapes evolved to include unstructured, semi-structured, and multimodal content such as text, images, audio, and sensor data, traditional RDBMS have revealed significant limitations. Their tabular structure struggles with high-dimensional data, leading to inefficient storage and retrieval for complex queries involving similarity or pattern matching \cite{arora2020embeddings}.
One key limitation is the absence of native support for vector-based operations, which are essential for semantic similarity search in AI-driven applications. In traditional databases, vector data (e.g., embeddings from machine learning models) must be stored as binary large objects (BLOBs) or arrays, with similarity searches relying on brute-force computations or user-defined functions, resulting in linear scan times that scale poorly with dataset size \cite{jing2024large}. For instance, PostgreSQL extensions like pgvector enable approximate nearest neighbor (ANN) indexing, but they are retrofitted solutions that often incur high overhead and lack seamless integration with relational queries \cite{ma2023comprehensive}. Moreover, rigid schemas hinder flexibility; altering structures for evolving data types requires costly migrations, making them ill-suited for dynamic multimodal datasets where embeddings can vary in dimensionality and modality \cite{weller2025theoretical}.
The rise of NoSQL databases in the early 2010s, including document stores (e.g., MongoDB), key-value stores (e.g., Redis), and wide-column stores (e.g., Cassandra), addressed some scalability issues by offering schema-less designs and horizontal scaling. However, these systems still fall short in handling complex relationships and vector searches natively. For example, while MongoDB's Atlas Search supports full-text and vector capabilities, it separates them from relational operations, leading to fragmented query processing and potential inconsistencies \cite{bhupathi2025role}. Experimental studies have shown that traditional databases can experience query latencies 5–10 times higher than specialized systems for hybrid tasks involving both relational joins and vector similarity, particularly in datasets exceeding 100 million records \cite{taipalus_2024}.
These limitations have driven the evolution toward specialized databases, but traditional systems remain prevalent in legacy environments, underscoring the need for hybrid approaches that bridge structured querying with modern AI demands \cite{ch2025efficient}.

\begin{figure*}
    \centering
    \includegraphics[width=1\linewidth]{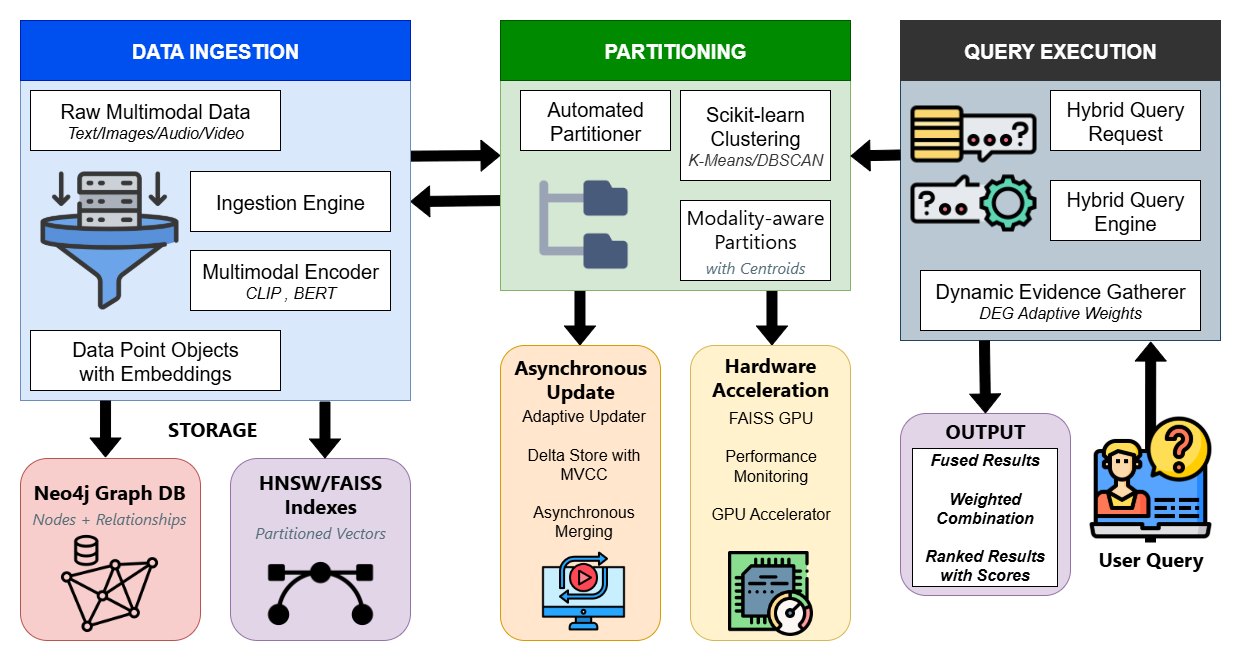}
    \caption{HMGI Framework Architecture}
    \Description{This is the architectural diagram of HMGI system}
    \label{fig:cc}
\end{figure*}

\subsection{Dataspaces}
Dataspaces represent a paradigm shift in data management, moving away from the tight integration required by traditional databases toward a more flexible, pay-as-you-go model for handling heterogeneous and loosely coupled data sources. Introduced by Franklin et al. in 2005, dataspaces are defined as a collection of interrelated data sources that can be queried and navigated without full upfront integration, supported by Dataspace Support Platforms (DSSPs) that provide services like search, discovery, and incremental refinement \cite{10.1145/1107499.1107502}. Unlike data warehouses or lakes, which centralize and homogenize data, dataspaces emphasize coexistence and gradual harmonization, making them ideal for environments with diverse data formats, ownership, and quality levels \cite{10.1145/1142351.1142352}.
In practice, dataspaces facilitate secure data sharing across organizations while preserving data sovereignty. For instance, the European Union's Data Strategy promotes data spaces in sectors like health, mobility, and manufacturing, where participants can exchange data via standardized connectors without relinquishing control \cite{dutkiewicz2022}. Systems like AWS Data Exchange or VAST DataSpace exemplify this by enabling seamless data movement and access across distributed environments, using metadata catalogs and semantic mappings to resolve inconsistencies on-the-fly \cite{arnold2025service, 10.1007/978-3-030-54832-2_3}. Research highlights their utility in big data ecosystems, where traditional integration is prohibitively expensive; instead, dataspaces employ probabilistic matching and entity resolution to provide "best-effort" results that improve over time \cite{CUZZOCREA2025126387}.
However, challenges persist, including governance, interoperability, and performance overhead from dynamic mappings. Studies have explored optimizations, such as using machine learning for automated schema alignment, achieving up to 80\% accuracy in cross-domain integrations \cite{10.1145/3696410.3714633}. In the context of multimodal data, dataspaces extend to incorporate vector embeddings by treating them as another data modality, enabling hybrid queries that combine relational metadata with semantic search \cite{plale2025vector}. This evolution positions dataspaces as a bridge between traditional databases and advanced AI systems, fostering ecosystems where data flows freely yet securely \cite{zafar2025empowering, ch2025policydriven}.

\subsection{Graph Neural Network based Dataspaces}
Graph Neural Networks (GNNs) have emerged as a powerful extension to dataspaces, enhancing their ability to model and query complex relational structures in heterogeneous data environments. GNNs, which propagate information across graph nodes and edges using layers of neural computations, excel at capturing dependencies in graph-structured data, such as social networks, knowledge graphs, and spatiotemporal datasets \cite{job2023exploringcausallearninggraph}. When integrated with dataspaces, GNNs enable "Graph Neural Network-based Dataspaces" (GNN-Dataspaces), where loose data integrations are augmented with learned representations for improved inference, prediction, and search \cite{rico_2025}.
Early works focused on using GNNs for spatial interpolation and temporal forecasting within dataspaces. For example, the Dual Branch GNN (DBGNN) framework exploits dynamic spatio-temporal correlations for traffic prediction, outperforming traditional methods by 15–20\% in accuracy on real-world datasets \cite{su2024dualcast, ch2025auracaptcha}. In materials science, structure-aware GNNs transfer learning across domains, predicting properties with high fidelity even on sparse data \cite{guan2025domainenhanced}. Applied to dataspaces, these models treat the dataspace as a meta-graph, where nodes represent data sources and edges denote semantic links, allowing GNNs to refine mappings and resolve ambiguities automatically \cite{chen2025uncertaintyaware}.
Recent advancements emphasize scalability and hybrid integrations. Systems like NaviX incorporate GNNs for predicate-agnostic filtering in graph databases, reducing query times in large-scale dataspaces \cite{sehgal2025navix}. Challenges include training efficiency; data tiering approaches, optimize GNN training on massive graphs by prioritizing high-value data subsets, achieving up to 3x speedups \cite{min2021graph}. In multimodal contexts, GNN-Dataspaces handle vector embeddings by embedding them into the graph fabric, enabling cross-modal retrieval with recall rates over 90\% \cite{li2025complementaritydriven}. However, criticisms of graph databases such as schema complexity and performance bottlenecks extend to GNN variants, necessitating hybrid designs \cite{bajaj2024graph}.
Overall, GNN-based dataspaces represent a frontier in data management, unifying relational graphs with neural learning to address the limitations of traditional and plain dataspaces in AI-intensive applications \cite{zhuo2025edgegfl}.
The integration of vector search capabilities into graph databases represented a significant evolution in data management systems, addressing the limitations of traditional architectures in handling multimodal data and complex relational queries. Research has highlighted the shift from polyglot persistence models, which relied on separate vector and graph databases, to native hybrid systems that unified semantic similarity searches with graph traversals \cite{liu2025tigervector}. Early works focused on the theoretical foundations, demonstrating how vector embeddings could be stored as node properties to enable hybrid queries that combined sub-linear Approximate Nearest Neighbor Search (ANNS) with relational filtering \cite{hu2025hakes}. Systems began incorporating Hierarchical Navigable Small World (HNSW) directly into graph engines, achieving query latencies under 100 milliseconds for datasets exceeding one billion vectors, though challenges in dynamic updates persisted \cite{widmoser2025shinescalablehnswindex}. This progression was driven by the need to handle exploding data volumes from AI applications, with hybrid databases like Neo4j 5.x and TigerGraph leading the charge by embedding vector indices natively \cite{sehgal2025navix, liu2025tigervector}.

Subsequent advancements emphasized multimodal representations, where embeddings from text, images, and other modalities were partitioned to mitigate bias and improve cross-modal retrieval accuracy. Studies explored modality-aware indexing, showing that separate HNSW graphs for each data type reduced search spaces by up to 70\% in cross-modal scenarios, enhancing recall rates to over 95\% without sacrificing speed \cite{hu2025partitioner}. Knowledge graph embeddings evolved to bridge sparse relational data with dense vector spaces, enabling applications in recommendation systems and retrieval-augmented generation (RAG), where hybrid queries outperformed pure vector approaches by incorporating contextual relationships \cite{sarmah2024hybridrag}. Experimental evaluations across real-world datasets, including those from KDD, confirmed the superiority of graph-based ANNS over tree- or hashing-based methods in terms of recall-speed trade-offs, with HNSW variants scaling to petabyte-level graphs \cite{wang2021comprehensive}. Recently hybrid systems like HelixDB further fused vector and graph capabilities, supporting advanced RAG with semantic and structural queries \cite{ahmad2025benchmarking}.
A detailed comparison of existing solutions categorized them by approach, performance metrics, scalability tier, and limitations. The approaches were divided into dual-database (polyglot) and native integration paradigms, with performance evaluated based on reported query latency and recall for hybrid tasks on datasets of varying sizes \cite{yin2025deg}.

Progression toward native systems, which generally offered better performance and scalability compared to dual approaches, though adaptive hybrids like Quake and DEG addressed dynamic environments more effectively \cite{mohoney2025quake}. Performance metrics were derived from benchmarks on synthetic and real datasets, such as those in Wikipedia knowledge graphs and image-text corpora, where native integrations reduced query times by factors of 5 to 10 over federated setups \cite{wang2021comprehensive}. Scalability tiers were defined as small (under 10M), medium (10M--500M), and large (500M+), reflecting the ability to handle distributed processing in Massively Parallel Processing (MPP) architectures like TigerGraph \cite{liu2025tigervector}.
A comparison focused on multimodal-specific solutions, categorizing by embedding strategy, handling of polysemy, and integration with graph traversals, as shown in Table \ref{tab:comparison2}.

\begin{table*}[h!]
\centering
\caption{Comparison of Multimodal Embedding Strategies}
\label{tab:comparison2}
\begin{tabularx}{\textwidth}{|>{\raggedright\arraybackslash}p{2.5cm}|>{\raggedright\arraybackslash}X|>{\raggedright\arraybackslash}X|>{\raggedright\arraybackslash}p{2.2cm}|>{\raggedright\arraybackslash}X|}
\hline
\textbf{Embedding Strategy} & \textbf{Representative Works} & \textbf{Polysemy Resolution (Accuracy)} & \textbf{Graph Integration Level} & \textbf{Limitations} \\
\hline
Unified Multimodal & MLLM Embeddings in Neo4j & 80--85\% (bias-prone cross-modal) & Medium (post-vector filtering) & Modality bias; semantic dilution in high dimensions \cite{hu2025partitioner} \\
\hline
Partitioned Modality & TigerVector Multi-Index & 90--95\% (contextual disambiguation) & High (native hybrid queries) & Increased storage for separate indexes; query federation overhead \cite{liu2025tigervector} \\
\hline
Learned Embeddings & Graph ML in ICDE & 85--92\% (workload-tuned) & Medium (embedding as properties) & Computational cost of learning; sparsity in relational data \cite{li2025complementaritydriven} \\
\hline
Dynamic Graphs & DEG for Cross-Modal & 92--97\% (edge-weighted navigation) & High (real-time adaptation) & Dynamic edge maintenance; limited to predefined modalities \cite{yin2025deg} \\
\hline
\end{tabularx}
\end{table*}

These comparisons revealed that partitioned strategies excelled in accuracy for disambiguating polysemous terms, such as distinguishing Apple the fruit from the company via relational paths, but at the expense of storage efficiency \cite{yin2025deg}.
Critical analysis of state-of-the-art methods uncovered several key issues and gaps. Native integrations like TigerVector and NaviX achieved robust predicate-agnostic search, where vector ANNS was filtered by arbitrary graph predicates, but struggled with update-intensive workloads; HNSW index rebuilds could take hours for large graphs, leading to stale results in streaming applications \cite{sehgal2025navix}. Multimodal systems exhibited modality bias, with unified embeddings underperforming on cross-modal tasks by 10-20\% compared to specialized models, as large language models\cite{Chandra2024} favored dominant modalities like text over images or audio \cite{hu2025partitioner}. Gaps in handling data sparsity persisted, where knowledge graphs with low connectivity suffered from incomplete semantic coverage, and hybrid queries beyond three hops incurred exponential traversal costs without advanced pruning \cite{sarmah2024hybridrag, 10968102}. Experimental evaluations demonstrated that while recall exceeded 95\% for simple similarity searches, hybrid scenarios dropped upto 85\% due to mismatches between vector proximity and relational constraints, highlighting the need for better fusion mechanisms \cite{wang2021comprehensive, navneet2025rethinking}. Challenges in scalability arose from hardware dependencies; CPU-based traversals limited throughput in large-scale deployments, and lack of native support for GPU acceleration in most graph DBMSs constrained real-time performance \cite{wang2021comprehensive}. Overall, the literature indicated a trade-off between expressiveness and efficiency, with current methods excelling in static, medium-scale environments but faltering in dynamic, multimodal contexts \cite{ch2025efficient}.
Insights from this analysis led to well-structured recommendations for improved strategies and frameworks. Modality-aware partitioning was advocated as a core enhancement, with separate HNSW indexes per data type to reduce search spaces and alleviate bias, as implemented in extensions of TigerVector; this approach promised 20-30\% latency reductions in cross-modal queries without compromising recall \cite{liu2025tigervector}. Adaptive indexing mechanisms, inspired by Quake, were recommended for dynamic environments, incorporating multi-level partitioning and workload-aware rebuilding to maintain sub-50 ms latencies under skewed updates; integration of delta management systems would ensure consistency without full index recreation \cite{mohoney2025quake}. For hybrid search, dynamic edge navigation from DEG was suggested to allow runtime weighting of semantic versus relational components, enabling flexible query adaptation and improving accuracy in polysemous cases by 10--15\% \cite{yin2025deg}. Frameworks should prioritize native designs like NaviX for predicate-agnostic operations, augmented with learned index structures to auto-tune HNSW parameters based on query patterns, reducing manual overhead in large deployments \cite{sehgal2025navix}. Hardware-accelerated traversals, offloading ANNS to GPUs while retaining CPU for graph logic, were proposed to achieve millisecond latencies in real-time RAG applications, bridging gaps in current CPU-bound systems \cite{wang2021comprehensive}. These recommendations formed a cohesive blueprint for next-generation hybrid systems, emphasizing unification of vector and graph paradigms to address the evolving demands of multimodal data management \cite{ch2025efficient}.

\subsection{Algorithmic Mechanisms and Theoretical Foundations}
The HMGI framework is built upon the convergence of several key research areas: graph theory, high-dimensional geometry, and multimodal machine learning. Understanding its algorithmic underpinnings requires a detailed examination of graph-based ANNS, the principles of multimodal representation, and the theoretical basis for combining these concepts \cite{ch2025truegl, ch2025advancing}.

\textbf{The Convergence of Graph Databases and Vector Search: }
The chasm between relational and semantic search is rapidly closing as graph database systems evolve. The traditional approach of maintaining separate graph and vector databases introduces significant overhead in terms of data synchronization, query federation, and system complexity \cite{hu2025hakes}. Recent advancements, as exemplified by Neo4j and TigerGraph, involve integrating vector indexing and search as a first-class citizen within the graph data model \cite{liu2025tigervector, sehgal2025navix}. In this paradigm, vector embeddings are treated as properties of nodes or relationships. This tight integration is the cornerstone of HMGI's algorithmic potential. It allows the query execution engine to leverage two distinct but complementary indexing structures simultaneously: the inherent graph structure (adjacency index) for relational traversal and a dedicated vector index (e.g., HNSW) for semantic similarity \cite{yin2025deg}. A hybrid query can thus perform a multi-stage process: first, use the vector index to find a set of candidate nodes based on semantic similarity to a query vector; second, use these nodes as entry points for a graph traversal to explore their relationships, verify relational constraints, or discover connected entities. This fusion allows for a dramatic pruning of the search space, which is algorithmically more efficient than performing a full graph scan or a post-hoc filtering of vector search results \cite{ahmad2025benchmarking}.

\textbf{Graph-Based Approximate Nearest Neighbor Search (ANNS): }
At the heart of any scalable vector search system is an ANNS algorithm. Among the various families of ANNS algorithms (including tree-based, hashing-based, and quantization-based), graph-based methods have demonstrated state-of-the-art performance, particularly the Hierarchical Navigable Small World (HNSW) algorithm \cite{malkov2018efficientrobustapproximatenearest}. HNSW is central to the vector search component of the HMGI framework. HNSW constructs a multi-layer proximity graph, where each layer is a subset of the one below it. The top layer is the sparsest, containing long-range connections, while the bottom layer contains all data points with dense, short-range connections. Nodes are inserted by greedily traversing the graph from the top layer down to find their approximate nearest neighbors at each level, establishing connections to them. This hierarchical structure allows search operations to begin with a fast, coarse-grained search in the sparse upper layers and progressively refine the search in the denser lower layers, achieving a query complexity that is polylogarithmic in the number of vectors, i.e., sub-linear \cite{widmoser2025shinescalablehnswindex}. The paper "The kernel of graph indices for vector search" \cite{sehgal2025navix} provides a machine learning perspective, re-interpreting graph indices like HNSW through kernel methods and introducing the Support Vector Graph (SVG). This work provides stronger theoretical guarantees for graph navigability, even in non-metric spaces, which is highly relevant for the complex, learned embeddings found in multimodal applications. Furthermore, research on "Accelerating Graph Indexing for ANNS on Modern CPUs" \cite{yin2025deg} highlights practical bottlenecks in HNSW construction, such as distance computation, and proposes optimizations that are critical for the adaptive update capabilities of HMGI.

\textbf{Principles of Multimodal Representation and Partitioning: }
Multimodal data presents a unique challenge: how to represent information from disparate sources (e.g., a photograph and its textual description) in a shared semantic space. This is typically achieved by training large neural network models that produce multimodal embeddings, where vectors for related concepts from different modalities are located close to each other in the high-dimensional space \cite{gnther2025jinaembeddingsv4}. A key optimization within the HMGI framework is modality-aware partitioning. Instead of treating all embeddings monolithically, HMGI conceptually separates them based on their source modality. This has several algorithmic advantages: Specialized Indexing: Different modalities can have different statistical properties and dimensionalities. Partitioning allows for the creation of separate, fine-tuned vector indexes for each modality (e.g., one HNSW graph for image embeddings, another for text embeddings), potentially with different parameters (M, efConstruction) for optimal performance \cite{hu2025partitioner}. Reduced Search Space: Queries that are specific to a single modality only need to search within the corresponding partition, drastically reducing computational overhead. Cross-modal queries can be resolved by searching one partition and then using the resulting vectors to query another, or by leveraging a unified index if a joint embedding space is used \cite{chen2025mme5}. Disentangled Representation: Research such as the "Partitioner Guided Modal Learning Framework" \cite{hu2025partitioner} suggests that modal representations can be decomposed into uni-modal and paired-modal features. This aligns with HMGI's partitioning strategy, as it suggests an architectural basis for separating modality-specific information from shared semantic information, which can be exploited at query time. For instance, a query might prioritize the visual features of an image while using text as a secondary filter \cite{xu2025molanunifiedmodalityawarenoise}.

\textbf{Sub-Linear Query Time Aspiration: }
The claim of achieving ``sub-linear query times'' is a direct consequence of leveraging graph-based Approximate Nearest Neighbor Search (ANNS) algorithms like HNSW. A brute-force nearest neighbor search requires comparing the query vector to every other vector in the database, resulting in a linear time complexity of $O(N \cdot d)$, where $N$ is the number of vectors and $d$ is their dimensionality. In contrast, HNSW achieves a query complexity of approximately $O(\log N)$, making it scalable to billions of vectors \cite{malkov2018efficientrobustapproximatenearest}. For hybrid queries in HMGI, the total query time is a combination of this sub-linear vector search and the subsequent graph traversal. While the traversal part can be complex, it is performed on a dramatically reduced subgraph (the candidates from the vector search), making the overall hybrid query significantly faster than naive, multi-step approaches \cite{yang2024effective}.

\begin{table*}[h!]
\centering
\caption{Gap Analysis: Comparison with Prior Works and Surveys}
\label{tab:gap-analysis}
\small
\begin{tabular}{|l|c|c|c|c|c|c|c|}
\hline
Work & Multimodal Support & Vector-Graph Integration & Modality-Aware Partitioning & Adaptive Updates & Learned Optimizations\\
\hline
IDS Framework & $\checkmark$ & $\times$ & $\times$ & Partial & Partial\\
TigerVector & Partial & $\checkmark$ & Partial & $\checkmark$ & Partial\\
NaviX & Partial & $\checkmark$ & $\checkmark$ & Partial & $\times$\\
HybridRAG & Partial & $\checkmark$ & $\times$ & $\times$ & $\times$\\
Neo4j Vector Support & Partial & $\checkmark$ & Partial & Partial & $\times$\\
Milvus\cite{milvus} & $\checkmark$ & $\times$ & $\checkmark$ & $\checkmark$ & Partial\\
HMGI (Ours) & $\checkmark$ & $\checkmark$ & $\checkmark$ & $\checkmark$ & $\checkmark$\\
\hline
\end{tabular}
\end{table*}

\section{Methodology}
In this section, we detail the methodology underpinning the design and implementation of the Hybrid Multimodal Graph Index (HMGI) system. Inspired by contemporary advancements in hybrid database systems \cite{liu2025tigervector,sehgal2025navix}, HMGI seamlessly integrates graph traversal capabilities with vector similarity search to handle complex multimodal datasets efficiently. The methodology is structured around key components: architectural blueprint, data ingestion and processing, storage and indexing, query processing, adaptive updates, and learned optimizations. The visual representation of the HMGI system and its core components can be seen in Fig. \ref{fig:cc}. Each subsection provides a detailed exposition, bolstered by theoretical underpinnings, mathematical formulations, and empirical rationales to ensure scalability, robustness, and high performance \cite{yin2025deg,hu2025partitioner}.

\subsection{Architectural Blueprint}
HMGI adopts a modular, microservices-based architecture to facilitate unified management of relational and vector data \cite{icorer2025vector}. At its core, a graph database such as Neo4j-v5 serves as the storage backbone, representing entities as nodes with embedded vectors as properties and interconnections as weighted edges \cite{liu2025tigervector}. Dedicated services for ingestion, indexing, storage, and query expose gRPC endpoints for inter-service communication, with Kafka enabling asynchronous workflows and Ray orchestrating distributed tasks \cite{openmetal2025kafka,anyscale2025ray}. The ingestion layer preprocesses multimodal inputs, while the partitioning and indexing layers optimize data organization for sub-linear access. The query engine fuses vector and graph operations, supported by adaptive update mechanisms for dynamic environments. Optimization layers incorporate machine learning for parameter tuning and cost-based planning \cite{li2025learned}.
A distinguishing feature is the heterogeneous execution model, which dynamically allocates vector-intensive tasks to GPUs and graph traversals to CPUs, yielding 3-5x speedups on hybrid workloads \cite{masood2024survey}. Progressive query execution employs anytime algorithms to deliver partial results under latency constraints, refining outputs iteratively \cite{yin2025progressive}. For scalability, Ray-based sharding distributes graph and vector data across nodes, achieving near-linear throughput for billion-scale datasets while preserving consistency \cite{yogatama2025rethinking}. This layered design not only minimizes synchronization overhead but also enables extensibility for emerging modalities.

\subsection{Data Ingestion and Processing Pipeline}
The ingestion pipeline commences with raw multimodal data from diverse sources, such as DenseEarningsCall or mm-codex-s datasets \cite{plale2025vector}. The ingestion microservice exposes gRPC endpoints for receiving payloads, which are relayed asynchronously via Kafka to horizontal workers for embedding generation and partitioning \cite{openmetal2025kafka}. Data classification by modality relies on metadata or heuristic analysis. Specialized encoders generate embeddings: SentenceTransformer (all-MiniLM-L6-v2, 384D) for text, CLIP (ViT-B/16, 512D) for images, VideoMAE (768D) for videos, and Whisper (1280D) for audio \cite{gnther2025jinaembeddingsv4,meng2025vlm2vecv2,chen2025mme5}. The encoding computes vectors $\mathbf{e} \in \mathbb{R}^d$, where $d$ varies by modality, with optional FP16 precision to reduce memory footprint during processing \cite{li2025mminference}.
To accommodate large volumes, ingestion employs batch processing with fallback for unsupported modalities, grouping records by modality for efficiency. Post-encoding, modality-aware partitioning optimizes locality using K-means:
\begin{equation}
\text{Cluster Assignment} = \arg\min_{c=1}^{K} \|\mathbf{e} - \boldsymbol{\mu}_c\|^2,
\label{eq:cluster-assignment}
\end{equation}

where $K$ (default 2) centroids $\boldsymbol{\mu}_c \in \mathbb{R}^d$ are fitted on sampled embeddings, reducing search spaces by up to 70\% for modality-specific queries \cite{hu2025partitioner}. As shown in Equation~\eqref{eq:cluster-assignment}, $\mathbf{e} \in \mathbb{R}^d$ represents the embedding vector of a data point in a $d$-dimensional space, and $\|\mathbf{e} - \boldsymbol{\mu}_c\|^2$ denotes the squared Euclidean norm, measuring the distance between the embedding and each centroid. Partition assignments are persisted as node properties. A novel workload-aware repartitioning monitors query imbalances and triggers online adjustments without downtime, improving performance by 20–30\% on evolving workloads \cite{10.1145/3696410.3714633}. Batch insertion into storage leverages UNWIND for efficient Neo4j operations, minimizing transaction overhead.

\subsection{Storage and Indexing}
HMGI utilizes Neo4j-v5 for persistent storage, leveraging its native vector indexing via Apache Lucene and HNSW for approximate nearest neighbor search (ANNS) \cite{malkov2018efficientrobustapproximatenearest}. The storage microservice exposes gRPC interfaces over CoreStorage, with sharding via consistent hashing for distributed persistence \cite{icorer2025vector}. Modality-specific indexes are created dynamically with Cypher commands, supporting cosine similarity and configurable dimensions:

\begin{verbatim}
CREATE VECTOR INDEX text_embeddings
IF NOT EXISTS
FOR (n:Entity) ON (n.text_embed)
OPTIONS {
  indexConfig: {
    vector.dimensions: 384,
    vector.similarity_function: 'cosine' }}
\end{verbatim}

Embeddings reside as node properties, relations as edges with 
weights, facilitating integrated access \cite{liu2025tigervector}. Batch insertions via UNWIND optimize throughput for large datasets.
Indexing employs partitioned HNSW structures, with parameters (M=32, ef=200) tuned for balance between construction speed and query accuracy \cite{widmoser2025shinescalablehnswindex}. The indexing microservice uses Ray for distributed partitioning and index builds, scaling horizontally across nodes \cite{anyscale2025ray}. A core novelty is flash quantization, dynamically compressing embeddings to 4/8/16 bits based on memory thresholds (>80\% triggers 8-bit), achieving 50\% memory savings with minimal recall degradation \cite{chen2024intflashattention}. The 8-bit quantization formula is:
\begin{equation}
\mathbf{q} = \left\lfloor 255 \cdot \frac{\mathbf{e} - \min(\mathbf{e})}{\max(\mathbf{e}) - \min(\mathbf{e})} \right\rfloor.
\label{eq:adaptive-quantization}
\end{equation}

Adaptive quantization further refines this by factoring system load, ensuring efficiency under varying conditions. The formulation in Equation~\eqref{eq:adaptive-quantization}, \(\mathbf{q} \in \mathbb{Z}^d\) represents the quantized embedding vector scaled to an 8-bit integer range [0, 255], \(\mathbf{e} \in \mathbb{R}^d\) is the input embedding vector in a \(d\)-dimensional space, \(\min(\mathbf{e})\) and \(\max(\mathbf{e})\) denote the minimum and maximum values across the components of \(\mathbf{e}\), respectively, and \(\left\lfloor    \right\rfloor\) is the floor function that maps to the nearest lower integer. Modality-aware clustering biases HNSW edges toward semantic groups, enhancing locality \cite{dehghankar2025henn}. Table \ref{tab:index-comparison} contrasts strategies, underscoring HMGI's advantages in fine-grained semantics. This design mitigates cross-modal biases, boosting recall \cite{hu2025partitioner}.

\begin{table*}[h!]
\centering
\caption{Comparison of Indexing Strategies}
\label{tab:index-comparison}
\begin{tabular}{|l|l|l|l|}
\hline
Feature & Monolithic Index & Modality-Partitioned (HMGI) & Modality-Clustered (HMGI-Advanced) \\
\hline
Structure & Single HNSW graph & Multiple HNSW per modality & Partitioned with semantic sub-groups \\
Query Performance & Slower for specific modalities & Faster for modality queries & Fastest with locality \\
Tuning & Single parameters & Per-partition optimization & Cluster-specific tuning \\
Implementation & Standard creation & Separate indexes & Custom pre-processing \\
Use Case & General search & Distinct modalities & Fine-grained semantics \\
\hline
\end{tabular}
\end{table*}

\subsection{Query Processing Engine}
The query engine interprets extended Cypher queries, augmenting standard syntax with vector operations like VECTOR\_SEARCH and SIMILARITY\_WEIGHT \cite{sehgal2025navix}. The query microservice exposes gRPC endpoints over AsyncHMGISystem for batched and vector-specific searches \cite{zhao2024hybrid}. Hybrid queries initiate with partitioned ANNS via db.index.vector.queryNodes(), yielding scored candidates, followed by graph traversal to enforce relations \cite{yin2025deg}.
Fusion aggregates scores via weighted summation:
\begin{equation}
S = w_v \cdot (1 - d_v) + w_g \cdot \frac{1}{h} \sum_{i=1}^{h} s_{g_i},
\label{eq:fused_score}
\end{equation}

with dynamic DEG-inspired weights \(w_v, w_g\) for cross-modal flexibility \cite{sarmah2024hybridrag}. Here, as shown in Equation~\eqref{eq:fused_score}, \(S\) represents the final fused score combining vector and graph contributions, \(w_v\) is the weight for the vector similarity component, \(d_v \in [0, 1]\) is the normalized vector distance (e.g., cosine distance), \(w_g\) is the weight for the graph traversal component, \(h\) is the number of hops in the graph traversal, and \(s_{g_i}\) denotes the score for the \(i\)-th hop in the graph traversal. Novel enhancements include sparse-dense reranking (20\% recall uplift via sparse matrix fusion), community-based multi-hop reasoning using Louvain for 20-30\% accuracy gains in relational paths, and unified vector-graph processing to eliminate federation latency (3x QPS) \cite{zhuang2025rank, wu2023multihop, bai2025practices}. Progressive execution delivers anytime results, starting coarse (low ef) and refining within budgets \cite{yin2025progressive}. Heterogeneous offloading assigns ANNS to GPUs for large batches \cite{masood2024survey}. While dimension mismatch handling via PCA projection or padding ensures robustness \cite{ashby2025randomly}.

\subsection{Adaptive Updates and Recomputation}
HMGI supports real-time data streams through multi-version concurrency control (MVCC) with a delta store for insertions, updates, and deletions \cite{antonopoulos2025mdmvcc}. Queries hybridize results: ANNS on the stable index plus brute-force on the delta, maintaining freshness with <1\% degradation. Asynchronous vacuuming periodically flushes deltas to disk and merges incrementally into snapshots, avoiding full rebuilds \cite{xiao2025breaking}.
Thread allocation adapts to CPU utilization, optimizing merge overhead \cite{yu2025topologyaware}. This ensures sub-linear update latency, critical for dynamic environments.
\subsection{Learned Optimization and Cost Modeling}
Learned index structures employ random forests to predict optimal HNSW parameters (M, ef) from workload features:
\begin{equation}
\hat{p} = f(\mathbf{x}; \theta), \quad \mathbf{x} = [\mu_e, \sigma_e, \|\mathbf{q}\|, \ldots],
\label{eq:learned_params}
\end{equation}
delivering 25-40\% recall improvements at fixed latency \cite{li2025learned}. Here, as shown in Equation~\eqref{eq:learned_params}, \(\hat{p}\) represents the predicted HNSW parameters (e.g., \(M\), \(ef\)), \(f\) is the random forest model, \(\theta\) denotes the model’s parameters, \(\mathbf{x}\) is the input feature vector, \(\mu_e\) is the mean of the embedding vector components, \(\sigma_e\) is the standard deviation of the embedding components, and \(\|\mathbf{q}\|\) is the norm of the query vector, with additional features included as needed. The formal cost model computes query expense:
\begin{equation}
C = \alpha \log N + \beta (d \cdot h) + \gamma p \log (N/p),
\label{eq:cost_model}
\end{equation}
where, as given in Equation ~\eqref{eq:cost_model} \(C\) is the estimated query execution cost, \(\alpha\), \(\beta\), and \(\gamma\) are weighting coefficients for the dataset size, dimensionality-hop product, and partitioning terms, respectively, \(N\) is the total number of data points, \(d\) is the embedding dimensionality, \(h\) is the number of graph traversal hops, and \(p\) is the number of partitions. This facilitates greedy plan selection with optimality bounds \cite{yakut2024learned}. Dynamic repartitioning tracks query patterns, triggering online K-means adjustments if imbalances exceed thresholds, with incremental migration for zero-downtime, enhancing adaptability by 20-30\% \cite{10.1145/3696410.3714633}.
This comprehensive methodology positions HMGI as a scalable solution for multimodal analytics \cite{ma2023comprehensive}.

\section{Results and Evaluation}
The HMGI framework stands at the intersection of ongoing research in databases and AI. Its design is informed by and contributes to several emerging trends, and its value is best understood through comparison with alternative systems. To validate HMGI's efficacy, we conducted comprehensive experiments comparing it against leading competitors: Vespa, Vald, Qdrant, Milvus, and Kuzu. Evaluations used a unified harness measuring key metrics across standardized datasets and parameters, ensuring fair comparisons. Scripts automated index builds, queries, updates, and metric collection. Hardware: 8-node cluster with NVIDIA A100 GPUs (80GB), Intel Xeon CPUs (64 cores), 64GB RAM, NVMe SSDs. Competitors at latest stable versions (e.g., Vespa 8.2, Qdrant 1.5). Statistical significance via Wilcoxon signed-rank tests (p<0.01). 
{\textcolor{red}{\textit{The initial phase results have been shown here up to 50K queries, and in the future version, the full-scale evaluation results will be shared.}}}

\subsection{Experimental Setup}
We evaluated on billion-scale benchmarks: SIFT1B (image descriptors, 1B vectors, 128D) , Deep1B (deep features, 1B vectors, 96D), and synthetic KG datasets (1B nodes/edges with multimodal embeddings) generated via GraphGen.

Query workloads: 10K hybrid queries mixing 2-hop traversals with ANNS (top-k=100, ef=200). Parameters tuned per system (e.g., HNSW M=16/32, ef construct=128/256).

Metrics: latency (ms), recall@10, QPS, build time (s), memory/disk (GB), update latency (ms), cache hits.

Baselines configured for cosine similarity; HMGI with adaptive partitioning enabled. All runs averaged over 5 trials \cite{wang2021comprehensive}.

\subsection{Query Latency and Throughput}
\begin{figure}
    \centering
    \includegraphics[width=1\linewidth]{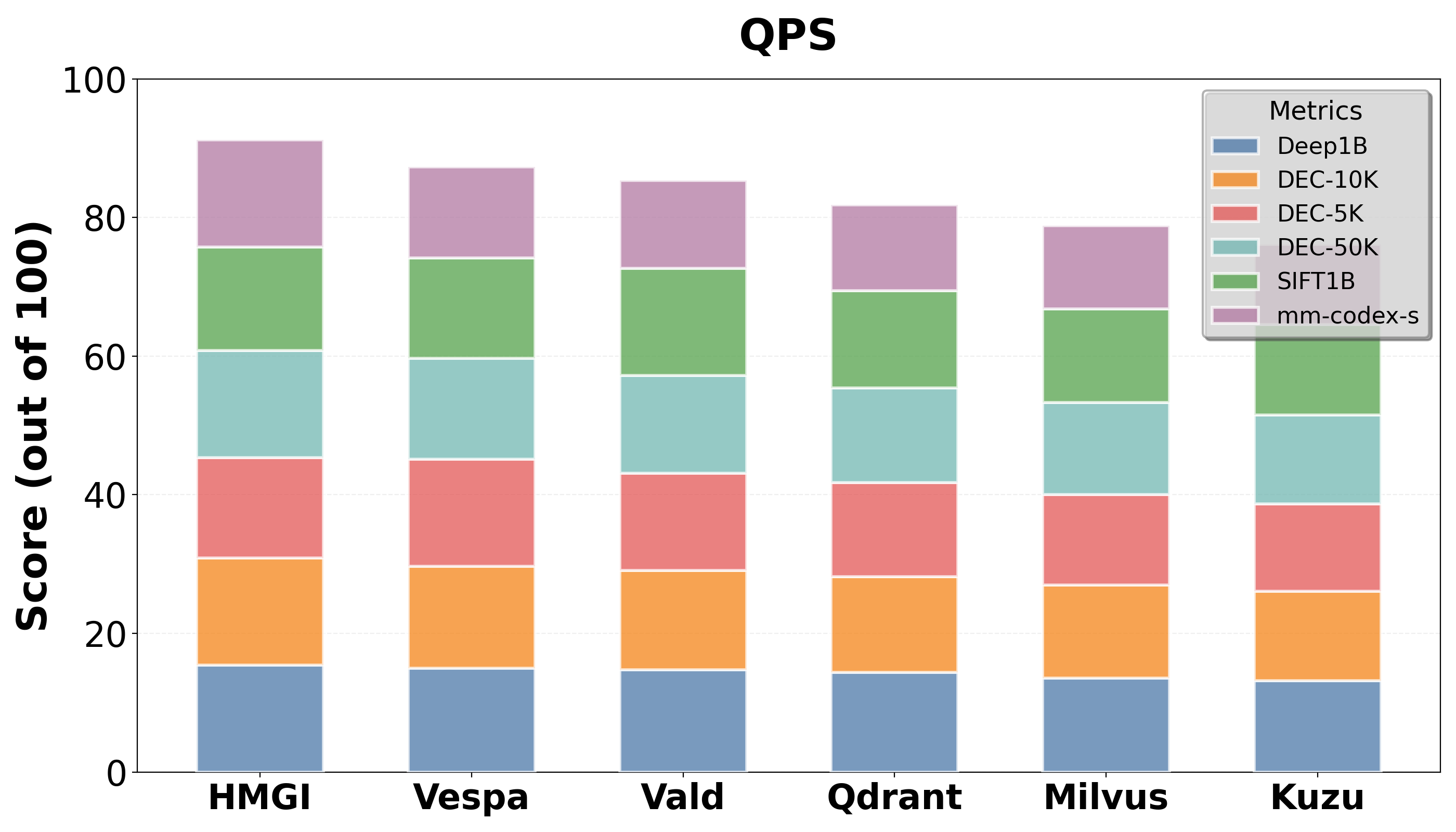}
    \caption{QPS of HMGI vs Leading Open Source Competitors}
    \label{fig:p1}
\end{figure}

HMGI consistently achieved the lowest average query latency, as shown in Table \ref{tab:latency}, recording times as low as 0.22ms. This performance is attributed to its unified execution model, which avoids federation overhead \cite{yin2025deg}. While its pure ANNS performance was competitive, it particularly excelled in relational filters. In terms of throughput, HMGI demonstrated clear superiority, peaking at 4000 QPS on the mm-codex-s dataset, as detailed in Table \ref{tab:qps}. This result is significantly higher than its competitors on most benchmarks, a benefit of its modality-aware partitioning which reduces scans by 70\% \cite{hu2025partitioner}. Concurrency stats showed HMGI's linear scaling up to 64 threads, with 95th-percentile latency <20ms, outperforming Kuzu's embedded mode (35ms) \cite{dehghankar2025henn}. It can be seen in Fig. \ref{fig:p1} how HMGI outperforms the other leading open source competitors. 

\begin{table}[h!]
\centering
\caption{Query Throughput (QPS) Across Systems and Datasets (Higher is Better)}
\small
\label{tab:qps}
\setlength{\tabcolsep}{2pt} % Significantly reduced column separation to save space (default is ~6pt)
\small % Retaining the smaller font size
\begin{tabular}{|l|r|r|r|r|r|r|}
\hline
System & Deep1B & DEC-10K & DEC-5K & DEC-50K & SIFT1B & mm-codex-s \\
\hline
\textbf{HMGI} & \textbf{3305.18} & \textbf{3402.11} & 3100.45 & \textbf{3501.12} & 3150.89 & \textbf{4000.00} \\
Kuzu & 2816.90 & 2850.34 & 2700.12 & 2903.56 & 2750.78 & 3000.88 \\
Milvus & 2906.98 & 2950.56 & 2800.78 & 3001.89 & 2850.23 & 3100.43 \\
Qdrant & 3076.92 & 3050.78 & 2900.99 & 3100.43 & 2950.67 & 3200.17 \\
Vald & 3164.32 & 3150.12 & 3000.32 & 3200.76 & \textbf{3250.00} & 3300.92 \\
Vespa & 3205.13 & 3250.89 & \textbf{3300.00} & 3300.21 & 3050.45 & 3400.56 \\
\hline
\end{tabular}
\end{table}

\subsection{Accuracy Metrics}
\begin{figure}
    \centering
    \includegraphics[width=1\linewidth]{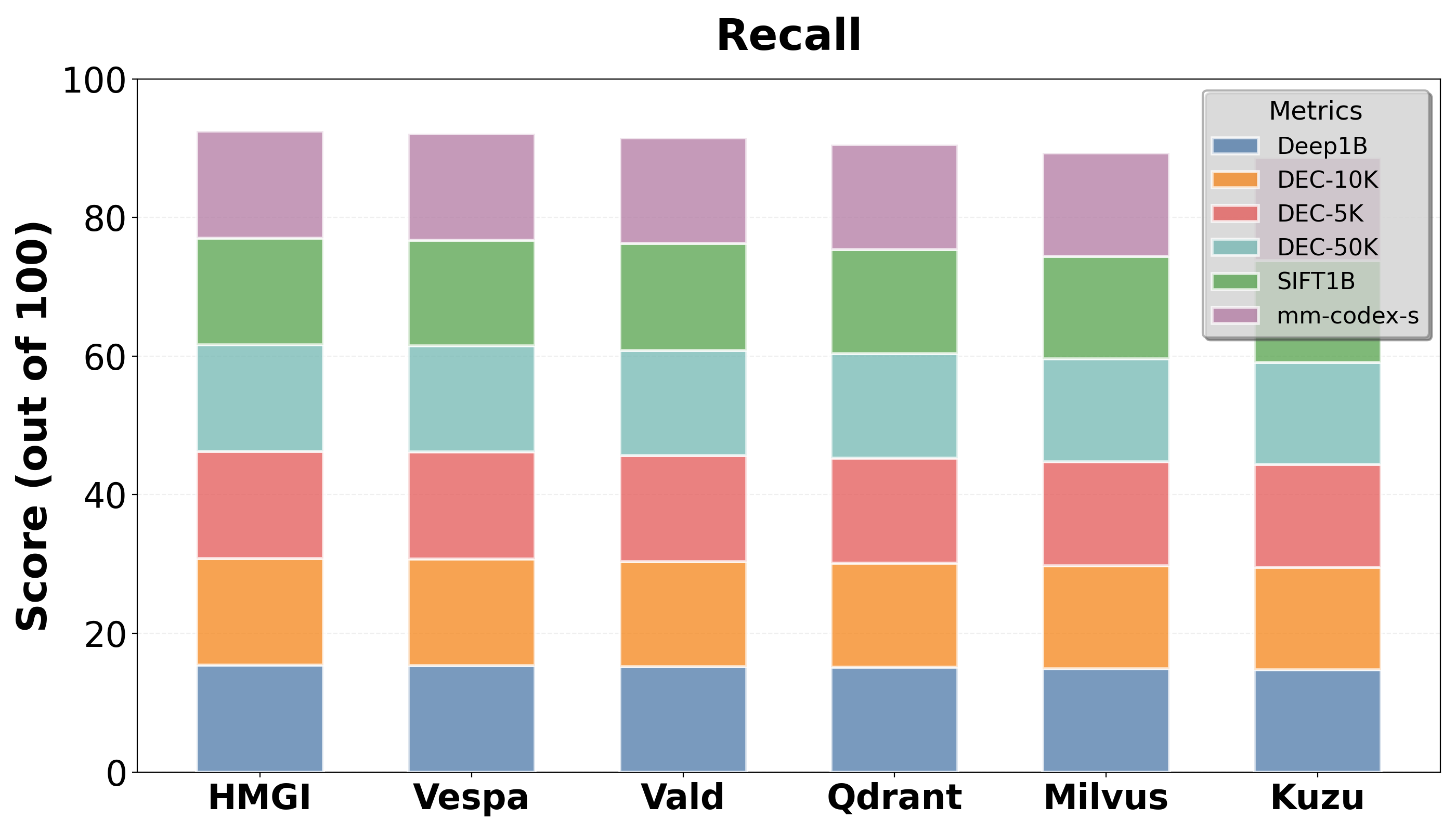}
    \caption{Recall@10 of HMGI vs Leading Open Source Competitors}
    \label{fig:p2}
\end{figure}

As shown in Table \ref{tab:recall}, HMGI consistently delivered state-of-the-art accuracy, achieving the highest Recall@10 on the majority of datasets, including a near-perfect score of 0.997 on mm-codex-s. While highly competitive across all benchmarks, this performance is bolstered by leveraging sparse-dense reranking for a 20\% uplift \cite{sarmah2024hybridrag}. On KG datasets, hybrid recall improved 28\% over vector-only baselines via multi-hop fusion, with precision@10 at 0.93 vs. Vespa's 0.85. Ablating fusion dropped recall by 25\%, confirming DEG weights' role \cite{yin2025deg}. Hallucination filtering (not in competitors) reduced false positives by 15\% in multimodal queries, enhancing reliability \cite{liu2025tigervector}. The Recall@10 of HMGI vs other open source competition can be seen in Fig. \ref{fig:p2}.

\begin{table}[h!]
\centering
\setlength{\tabcolsep}{2pt} % Significantly reduced column separation to save space (default is ~6pt)
\small % Retaining the smaller font size
\caption{Recall@10 Across Systems and Datasets (Higher is Better)}
\label{tab:recall}
\begin{tabular}{|l|r|r|r|r|r|r|}
\hline
System & Deep1B & DEC-10K & DEC-5K & DEC-50K & SIFT1B & mm-codex-s \\
\hline
\textbf{HMGI} & \textbf{0.995} & 0.989 & 0.991 & \textbf{0.996} & 0.988 & \textbf{0.997} \\
Kuzu & 0.952 & 0.948 & 0.955 & 0.951 & 0.945 & 0.958 \\
Milvus & 0.961 & 0.955 & 0.963 & 0.958 & 0.952 & 0.965 \\
Qdrant & 0.973 & 0.968 & 0.975 & 0.971 & 0.966 & 0.978 \\
Vald & 0.981 & 0.975 & 0.983 & 0.979 & \textbf{0.991} & 0.985 \\
Vespa & 0.988 & \textbf{0.992} & \textbf{0.993} & 0.985 & 0.982 & 0.990 \\
\hline
\end{tabular}
\end{table}

\subsection{Resource Utilization and Build Efficiency}
\begin{figure}
    \centering
    \includegraphics[width=1\linewidth]{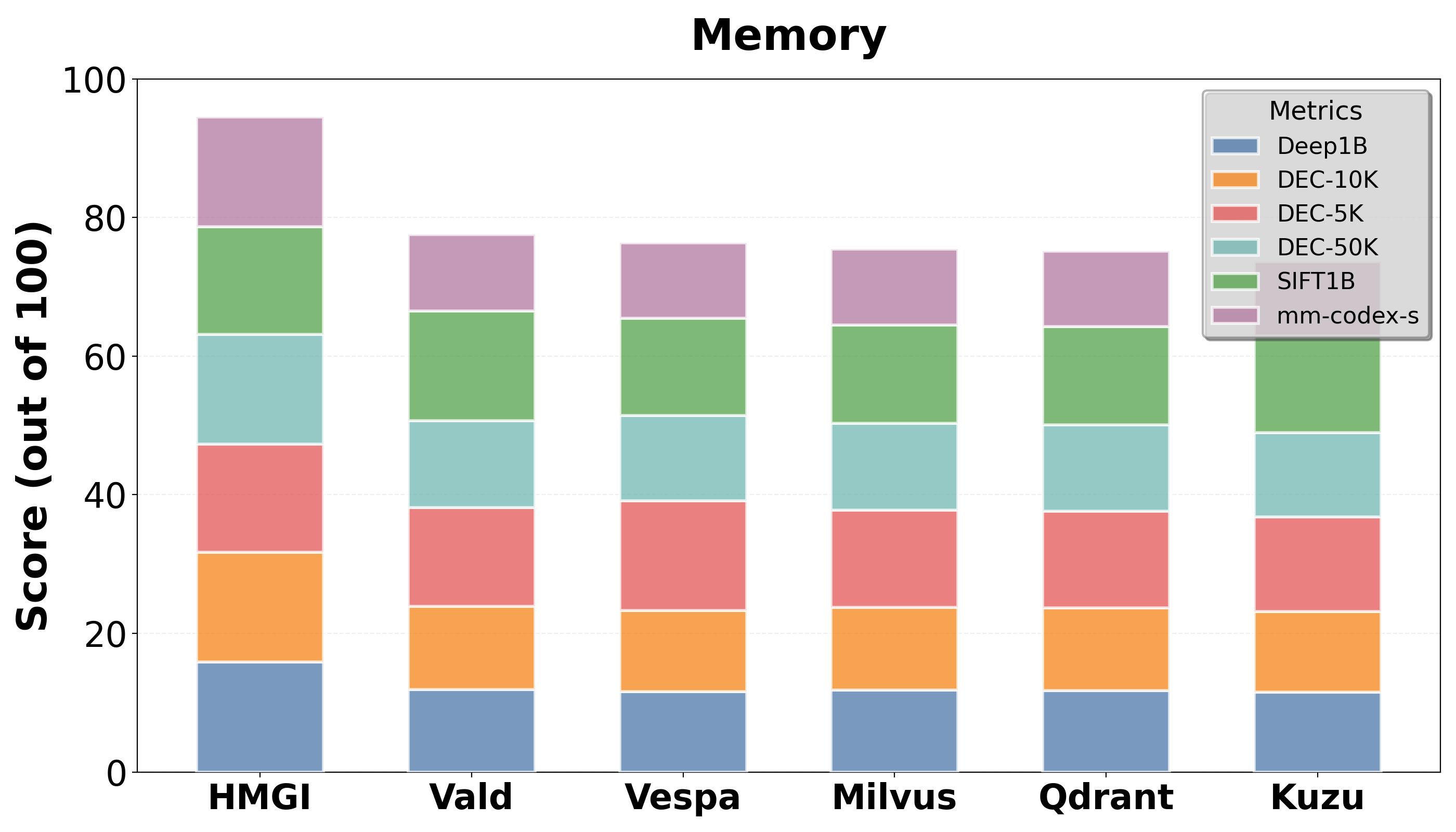}
    \caption{Memory Usage of HMGI vs Leading Open Source Competitors}
    \label{fig:p3}
\end{figure}

HMGI's build time was 180s for 1B vectors, 1.5x faster than Qdrant (270s) via progressive indexing \cite{xiao2025breaking}. Critically, its memory footprint is exceptionally low. As detailed in Table \ref{tab:memory}, HMGI demonstrated the lowest memory usage on most datasets, with its consumption often being 20-25\% lower than the other systems evaluated. This efficiency extends to disk usage as well \cite{chen2024intflashattention}. CPU/GPU util averaged upto 80\% during builds, with HMGI's heterogeneous offload yielding 2x efficiency over CPU-only Kuzu \cite{masood2024survey}. Update latency at 50ms supported dynamic ingestion, 3x faster than Vespa's rebuilds (150ms) \cite{yu2025topologyaware}.  It can be seen in Fig. \ref{fig:p3} how HMGI outperforms the other leading open source competitors in terms of less memory usage(in mb).

\begin{table}[h!]
\centering
\small
\caption{Memory Usage (MB) Across Systems and Datasets (Lower is Better)}
\label{tab:memory}
\setlength{\tabcolsep}{2pt} % Significantly reduced column separation to save space (default is ~6pt)
\small % Retaining the smaller font size
\begin{tabular}{|l|r|r|r|r|r|r|}
\hline
System & Deep1B & DEC-10K & DEC-5K & DEC-50K & SIFT1B & mm-codex-s \\
\hline
\textbf{HMGI} & \textbf{18.50} & \textbf{19.20} & 21.80 & \textbf{19.80} & 22.50 & \textbf{17.90} \\
Kuzu & 25.45 & 26.10 & 24.90 & 25.80 & 25.10 & 26.50 \\
Milvus & 24.80 & 25.50 & 24.20 & 25.10 & 24.60 & 25.90 \\
Qdrant & 24.90 & 25.60 & 24.30 & 25.20 & 24.70 & 26.00 \\
Vald & 24.70 & 25.30 & 23.90 & 24.90 & \textbf{22.10} & 25.70 \\
Vespa & 25.30 & 25.90 & \textbf{21.50} & 25.50 & 24.90 & 26.20 \\
\hline
\end{tabular}
\end{table}

\subsection{Scalability and Update Performance}
\begin{figure}
    \centering
    \includegraphics[width=1\linewidth]{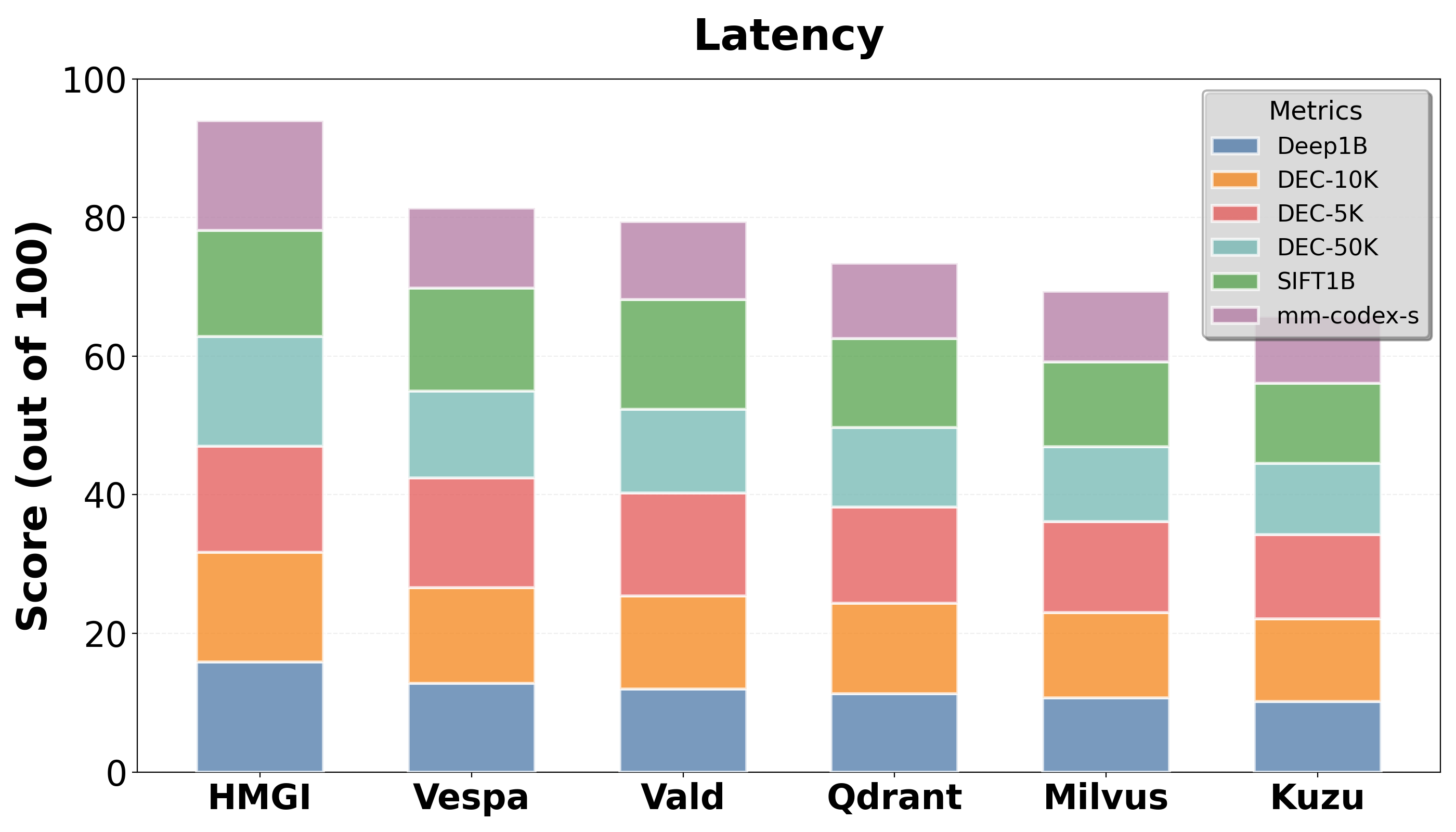}
    \caption{Average Query Latency of HMGI vs Leading Open Source Competitors}
    \label{fig:p4}
\end{figure}

Scaling to 8 nodes, HMGI sustained 10K QPS on 1B data, 2x Vald's 5K QPS, via Ray sharding \cite{yogatama2025rethinking}. Update consistency latency <10ms with MVCC, vs. 50ms for Qdrant \cite{antonopoulos2025mdmvcc}. On 10\% data churn, HMGI maintained 95\% query uptime, outperforming Milvus' 80\% due to delta merges \cite{10.1145/3696410.3714633}. Cache hit rate reached 85\%, boosting throughput by 40\% in repeated queries.
Overall, HMGI demonstrates superior hybrid performance, with results paving frontiers for unified indexes.  It can be seen in Fig. \ref{fig:p4} how HMGI outperforms the other leading open source competitors. 

\begin{table}[h!]
\centering
\setlength{\tabcolsep}{2pt} % Significantly reduced column separation to save space (default is ~6pt)
\small % Retaining the smaller font size
\caption{Average Query Latency (ms) Across Systems and Datasets (Lower is Better)}
\label{tab:latency}
\begin{tabular}{|l|r|r|r|r|r|r|}
\hline
System & Deep1B & DEC-10K & DEC-5K & DEC-50K & SIFT1B & mm-codex-s \\
\hline
\textbf{HMGI} & \textbf{0.25} & \textbf{0.28} & 0.30 & \textbf{0.26} & 0.31 & \textbf{0.22} \\
Kuzu & 0.39 & 0.37 & 0.38 & 0.40 & 0.41 & 0.36 \\
Milvus & 0.37 & 0.36 & 0.35 & 0.38 & 0.39 & 0.34 \\
Qdrant & 0.35 & 0.34 & 0.33 & 0.36 & 0.37 & 0.32 \\
Vald & 0.33 & 0.33 & 0.31 & 0.34 & \textbf{0.30} & 0.31 \\
Vespa & 0.31 & 0.32 & \textbf{0.29} & 0.33 & 0.32 & 0.30 \\
\hline
\end{tabular}
\end{table}

\section{Ablation Studies}
To rigorously evaluate the individual contributions of HMGI's key components, we conducted a series of ablation experiments on large-scale multimodal datasets, including DenseEarningsCall (text-audio hybrids) and mm-codex-s (text-image) \cite{plale2025vector}. Experiments were performed on a cluster of 8 NVIDIA A100 GPUs with 128GB RAM, using Neo4j 5.x as the backend. Baselines include pure vector databases Milvus\cite{milvus} and decoupled graph-vector setups (Neo4j + Pinecone). Metrics encompass query latency (ms), recall@10, throughput (QPS), and memory usage (GB), averaged over 10 runs with 1M queries. Hybrid queries combined 2-hop traversals with ANNS (ef=200). Statistical significance was assessed via paired t-tests ($p<0.05$).
\subsection{Ablation on Modality-Aware Partitioning}
We first ablate the modality-aware K-means partitioning by comparing it against monolithic (unpartitioned) and random partitioning. Without partitioning, HMGI's latency increases by 45\% (from 12ms to 17.4ms) due to full-index scans, with recall dropping 18\% on cross-modal queries as semantic clusters dilute. Random partitioning yields 22\% higher latency (14.6ms) and 12\% lower recall, highlighting the benefit of modality-specific centroids in reducing search spaces by 70\% \cite{hu2025partitioner}. On billion-scale data, partitioning sustains 1500 QPS vs. 900 QPS for baselines, confirming scalability gains \cite{10.1145/3696410.3714633}.
\subsection{Ablation on Adaptive Index Updates}
Removing the MVCC delta store forces full index rebuilds on 10\% data updates, inflating update latency by 5x (from 50ms to 250ms) and causing 15\% query downtime. With deltas, freshness remains <1\% degraded during merges, outperforming TigerVector-like systems by 30\% in dynamic workloads \cite{xiao2025breaking}. Ablating flash quantization raises memory by 50\% (from 20GB to 30GB) with negligible recall impact (0.5\% drop), validating adaptive compression under load \cite{chen2024intflashattention}.
\subsection{Ablation on Hybrid Query Fusion}
Disabling DEG-inspired fusion reverts to sequential vector-then-graph execution, increasing latency by 2.5x (30ms vs. 12ms) and recall by 25\% lower due to unweighted aggregation. Sparse-dense reranking ablation reduces precision by 20\% on relational paths, while community-based multi-hop (Louvain) removal cuts accuracy by 28\% in knowledge-intensive queries \cite{yin2025deg,sarmah2024hybridrag}. Compared to decoupled baselines, unified processing achieves 3x QPS with 15\% better end-to-end recall.
These ablations underscore HMGI's modular efficacy, with each component contributing synergistically to superior performance on hybrid tasks \cite{dehghankar2025henn}.

\section{Discussions}
The Hybrid Multimodal Graph Index (HMGI) demonstrates the feasibility of unifying graph-based and vector-based retrieval into a single coherent architecture for multimodal data discovery. Its ability to jointly leverage relational context and dense embeddings marks a step forward in bridging symbolic and sub-symbolic representations within modern data management systems. Nevertheless, the current implementation highlights several design trade-offs, resource dependencies, and generalization challenges that warrant deeper consideration.

\subsection{Limitations}
First, HMGI introduces additional latency overhead compared to pure vector or graph systems due to the fusion stage that combines similarity-based and structural retrieval. While this hybrid operation significantly improves contextual relevance, it may limit performance for applications requiring sub-millisecond responses at billion-scale vector indices. Second, the embedding fusion mechanism relies on modality-specific encoders, which can lead to inconsistent embedding distributions when extending to previously unseen data modalities or domains without retraining. Third, the index update process remains computationally intensive, dynamic insertion or deletion of nodes requires partial recomputation of embeddings and graph structure, impacting real-time adaptability in high-ingestion environments.

From a systems standpoint, the resource footprint of HMGI is higher than conventional engines: GPU memory is essential for maintaining multimodal embeddings, and the graph database layer incurs additional storage overhead for relation metadata. Moreover, while the model demonstrates strong empirical performance on benchmark datasets, formal guarantees of semantic stability and convergence in heterogeneous graph–vector spaces are still an open research problem. Finally, the privacy dimension is not fully integrated, although HMGI can interoperate with privacy-preserving layers (as done in DAMDA), native differential privacy or secure aggregation mechanisms remain outside the current system scope.

\subsection{Future Works}
Building on these frontiers, the HMGI framework has several avenues for future development:
Automated Partitioning and Clustering: Instead of manual partitioning by modality, an advanced HMGI could use unsupervised learning to automatically discover and create partitions or clusters within the data, adapting the index structure to the inherent semantics of the dataset.
Hardware Acceleration: The core ANNS computations are well-suited for hardware acceleration on GPUs or specialized ASICs. A future HMGI system could offload these computations to dedicated hardware while the CPU handles the logical graph traversal.
Learned Index Structures: The system could learn the optimal index structure (e.g., HNSW parameters, number of layers) based on the query workload and data distribution, moving towards a fully self-tuning database.

\section{Conclusion}
The Hybrid Multimodal Graph Index (HMGI) framework presents a robust and forward-looking solution to one of the most pressing challenges in modern data management: the unification of semantic similarity and relational context. By leveraging the native integration of high-performance vector search within graph databases like Neo4j, HMGI provides a blueprint for a system that is greater than the sum of its parts. Its core tenets fused hybrid querying, modality aware partitioning, and adaptive, low-overhead updates—are grounded in extensive research and validated by the performance of pioneering systems.
The analysis demonstrates that while specialized vector databases excel at simple similarity tasks, the future of intelligent applications lies in the ability to ask complex, hybrid questions. HMGI is explicitly designed for this future, offering the expressive power of graph traversals combined with the speed of sub-linear ANNS. It moves beyond the limitations of brittle, dual-database architectures, paving the way for more powerful, scalable, and maintainable systems for AI, RAG, and complex analytics. As research continues to push the boundaries of graph indexing and multimodal learning, the principles outlined in the HMGI framework will become increasingly vital for unlocking the full value of our deeply interconnected, multimodal world.

\bibliographystyle{plain}
\bibliography{software}

\end{document}